\documentclass[12pt]{article}
\usepackage{amsfonts,graphicx}

\def\d{{\rm d}}
\def\mi{{\rm i}}

\def\G{\mathop{\Gamma}\nolimits}
\def\l{{\lambda}}
\def\Re{\mathop{\rm Re\,}\nolimits}

\def\Tr{\mathop{\rm Tr\,}\nolimits}
\def\e{\mathop{\rm e}\nolimits}

\def\WKB{{\rm WKB}}
\def\hf{{\textstyle{1 \over 2}}}

\def\qt{{\textstyle{1 \over 4}}}

\def\N2{{\textstyle{\displaystyle N \over \displaystyle 2 \strut}}}
\def\defi{\stackrel{\rm def}{=}}
\def\si{\!\!\! &}
\def\se{& \!\!\!}
\def\ni{\noindent}

\newcommand{\beq}{\begin{equation}}
\newcommand{\eeq}{\end{equation}}
\newcommand{\bea}{\begin{eqnarray}}
\newcommand{\eea}{\end{eqnarray}}
\newcommand{\ba}{\begin{array}}
\newcommand{\ea}{\end{array}}

\title{Zeta-regularisation for exact-WKB resolution of a general 1D Schr\"odinger equation}

\author{{\bf Andr\'e Voros}\\
CEA-Institut de Physique Th\'eorique de Saclay\\
F-91191 Gif-sur-Yvette Cedex (France)\\
E-mail: \tt andre.voros@cea.fr}

\begin{document}

\maketitle

\begin{abstract}
We review an exact analytical resolution method for general one-dimensional (1D) 
quantal anharmonic oscillators: stationary Schr\"odinger equations with polynomial
potentials. It is an exact form of WKB treatment involving spectral (usual)
vs ``classical" (newer) zeta-regula\-risations in parallel. 
The central results are a set of Bohr--Sommerfeld-like but exact quantisation
conditions, directly drawn from Wronskian identities,
and appearing to extend Bethe-Ansatz formulae of integrable systems.
Such exact quantisation conditions do not just select the eigenvalues;
some evaluate the spectral determinants, and others the wavefunctions, 
for the spectral parameter in general position.
\end{abstract}

It is a great honour and pleasure to write in this volume for Professor Dowker 
as a tribute to his prominent contributions to mathematical physics. 
Because of our common interests toward functional (i.e., zeta-regularised) determinants 
in spectral theory, we stay in this area. However, instead of the more usual 
applications in geometry (e.g., \cite{VZ,D}), quantum fields (e.g., \cite{EZ,O}) 
and analytic number theory (e.g., \cite{DK,VA}),
here we propose a foray into analysis, with an exact treatment for Sturm--Liouville equations
--- or 1D stationary Schr\"odinger equations in quantum mechanics.

We specifically review our exact-WKB resolution of that problem, 
trying to draw the main lines in a global perspective, 
which is blurred by many technicalities in our earlier works.
Therefore we make no attempt at completeness here and may discard even important details
for which we refer to earlier presentations (\cite{V5,V7,V9,VK} on the technical side;
\cite{V8,V0} on the survey side).
Likewise, the bibliography here is partial and largely meant to refer to those in our earlier
works, most developed in \cite{V7,V0,VK}.
Inversely, two points previously scattered among several articles are revisited here 
in greater detail: the classical side of zeta-regularisation in Sect.~2,
and the proof of the basic exact-WKB identities (\ref{BID}) in Sect.~3 and the Appendix.

\section{General setting}

We review an analytical resolution method, through an exact-WKB approach assisted by zeta-regularisation,
for the stationary 1D Schr\"odinger equation (or Sturm--Liouville problem)
\beq
\label{SSE}
-\psi'' (q) + \bigl[ V(q)+\l \bigr] \, \psi (q)=0 ,
\qquad \Bigl( \, ' \equiv \frac{\d}{\d q} \Bigr),
\eeq
with $V(q)$ (the potential) a {\sl polynomial\/} function, here normalised as
\beq
V(q)= {+ \, q^N + \, v_1 \,q^{N-1} + \cdots + \, v_{N-1} \,q} \qquad (N \ge 1) .
\eeq
Key parameters will be
\beq
\label{PAR}
N \mbox{ (degree)}, \quad \mu \defi \frac{1}{2}+\frac{1}{N} \mbox{ (order)}, 
\quad \varphi \defi \frac{4\pi}{N+2} \mbox{ (symmetry angle)},
\eeq
plus the ``number of conjugates" $L$ to be determined in (\ref{NCJ}).
Often we will simplify results by excluding $N=2$, 
which is most singular as special case within the general-$N$ theory. 

Other quantum-mechanical notations will also help: 
$\hat H$ (Schr\"odinger operator), $E$ (energy variable), condensing (\ref{SSE}) to
\beq
\label{SEE}
\hat H \psi = E \psi, \qquad {\rm with} \quad \hat H \defi -{\d^2 \over \d q^2} + V(q), \quad E \equiv -\l , 
\eeq
but this continues to mean the bare differential equation (\ref{SSE}):
$E$ is not restricted a priori to eigenvalues, nor $\psi$ to some operator domain.

Being motivated by self-adjoint bound-state calculations in quantum mechanics,
we envision both $q$ and $V$ as real, or 
$\vec v \defi {(v_1,\ldots,v_{N-1}) \in {\mathbb R}^{N-1}}$.
For such spectral problems the best setting is a real half-line, 
say $q \in [0,+\infty)$, with ultimately a Dirichlet or Neumann boundary condition at the endpoint $q=0$.
Still, our exact resolution will involve 

- the whole complex $q$-plane;

- all ``conjugate equations" of the form (\ref{SSE}) but complex,
that restore (\ref{SSE}) under some analytic dilation $q \mapsto \e^{\mi \theta}q$;
\cite[\S 2.7]{S} there are finitely many, precisely
\beq
\label{NCJ}
L = \left\{ \matrix{ 
N+2 \hfill & \mbox{in general} \hfill \cr
\hf N + 1 & \mbox{for an \sl even polynomial }V(q)} \right\}
\mbox{ distinct conjugates} :
\eeq
\beq
\label{CJE}
V^{[\ell]}(q) \defi \e^{-\mi\ell\varphi} V(\e^{-\mi\ell\varphi/2}q), 
\ \l ^{[\ell]} \defi \e^{-\mi\ell\varphi} \l
\quad \mbox{for } \ell=0,1,\ldots,L-1 \pmod L, 
\eeq
with the angle $\varphi = 4\pi/(N+2)$ as announced in (\ref{PAR}).

Traditionally, the ordinary differential equation (\ref{SSE}) only admitted asymptotic 
(hence approximate) solutions in analytic form:  
{\sl WKB expressions\/}, valid for $q$ or $\l $ large.
Then, parts of WKB theory have been turned fully exact \cite{DI,BB,S,V1,AKT2}.
What we will outline here is a {\sl global exact-WKB\/} treatment for (\ref{SSE}), 
which just falls short of self-sufficiency:
curiously, it still needs explicit inputs of prior asymptotic-WKB results; 
hence we review these first.

\section{Asymptotic ingredients}

\subsection{Known spectral properties}

To streamline this part, we phrase it for {\sl even polynomials\/} $V(q)$.
(The non-even case adds minor, but tedious, complications. \cite{V7})
Then, the Schr\"odinger operator $\hat H$ over all of $\mathbb R$,
with $V(q) \uparrow +\infty$ at both ends $q \to \pm\infty$, 
has a discrete countable $E$-spectrum
${\mathcal E} \defi \{ E_k \}_{k=0,1,\ldots}$ with $E_k \uparrow +\infty$.
By {\sl parity-symmetry\/}, moreover, 
${\mathcal E}^+ \defi \{ E_k \}_{k \ \rm even}$ is the Neumann spectrum
for (\ref{SEE}) over the half-line $q \in [0,+\infty)$,
while ${\mathcal E}^- \defi \{ E_k \}_{k \ \rm odd}$ is the Dirichlet spectrum.

\subsubsection{Semiclassical asymptotics}
The spectral semiclassical regime is $k \to +\infty$, 
for which the eigenvalues $E_k$ asymptotically obey a Bohr--Sommerfeld condition
shaped by the classical Hamiltonian $p^2+V(q)$ over the phase space ${\mathbb R}^2$, 
as
\cite[\S 10.5]{BO}
\beq
\label{BSC}
\oint_{p^2+V(q)=E} \frac{p \, \d q}{2\pi} \Biggm\vert_{E=E_k} \sim k+\hf 
\qquad \mbox{for integer } k \to +\infty ;
\eeq
the leading power behaviour in $E \to +\infty$ is
\beq
\oint_{p^2+V(q)=E} \frac{p \, \d q}{2\pi} \sim b_\mu E^\mu \quad \mbox{with } \mu = \frac{1}{2}+\frac{1}{N}
\quad (\mbox{and } b_\mu = \oint_{p^2+q^N=1} \frac{p \, \d q}{2\pi} ) ,
\eeq
introducing an essential parameter, the {\sl order\/} $\mu \in (\hf,\frac{3}{2}]$. 
However, a stronger result needs to be known: that the {\sl complete\/} Bohr--Sommerfeld formula, 
which incorporates the semiclassical corrections to (\ref{BSC}) to all orders 
\cite[\S 10.7]{BO}, 
admits a complete $E \to +\infty$ expansion as a {\sl descending power series\/},
\beq
\label{SCE}
\sum_{\alpha \in \mathcal A} b_\alpha {E_k}^\alpha \sim k + \hf \quad
\mbox{for integer } k \to + \infty , \quad \textstyle 
{\mathcal A} = \bigl\{ \mu,\ \mu-{1 \over N},\ \mu-{2 \over N}, \ldots \bigr\} 
\eeq
(each $b_\alpha$ is a polynomial expression in the $\{v_j\}_{j \le (\mu-\alpha)N}$).

\subsubsection{Spectral functions}

The form  (\ref{SCE}) implies the following statements, issued for real $\l > -\min V$ initially
(other $\l $ are to be reached by analytical continuation).

- The pair of series
\beq
\label{SQZ}
Z^\pm(s,\l ) \defi \! \sum_{k \ {\rm even \atop odd}}
(E_k + \l ) ^{-s} \qquad \mbox{ for } \Re s > \mu 
\eeq
converge, defining analytic functions $Z^\pm$: {\sl (generalised) spectral zeta-functions\/}.

- The summations in (\ref{SQZ}) can be handled by Euler--Maclaurin formulae,
whereby $Z^\pm$ continue to meromorphic functions in the whole complex $s$-plane, 
{\sl pole-free (i.e., regular) at\/} $s=0$. \cite[\S 1.2]{V5} 

Besides, (term by term) all sums of the form (\ref{SQZ}) obey the general identity
\beq
\label{ZFR}
\partial_\l Z(s,\l ) \equiv -sZ(s+1,\l ) .
\eeq

Next, the regularity of $Z^\pm$ at $s=0$ allows to specify {\sl spectral determinants\/}, not just formally as
\beq
\label{FQD}
D^\pm(\l ) = \mbox{``} \!\prod\limits_{k \ {\rm even \atop odd}} 
(\l \!+\! E_k) \mbox{ "} \qquad \mbox{(severely divergent)},
\eeq
but rigorously, as well-defined entire functions of $\l $, through
\beq
\label{SQD}
\log D^\pm(\l ) \defi - \partial_s Z^\pm(s,\l )_{s=0}
\qquad  (\mbox{\sl zeta-regularisation}) .
\eeq

In (\ref{SQD}), the outputs of Euler--Maclaurin summation are limit formulae 
which we call {\sl structure formulae\/} (akin to Hadamard products, but with {\sl no free constants\/} left),
\bea
\label{EML}
\log D^\pm(\l ) \si=\se
\lim_{K \to + \infty}  \left\{ \sum_ {k<K} \log (E_{k} + \l )
+ \hf \, \log (E_{K} \!+\! \l ) \right. \nonumber \\[-10pt]
\\[-10pt]
\si\se \qquad \qquad {} - \hf \sum_{ \{\alpha >0 \} }
\left. \vphantom{\sum_K}
 b_{\alpha} (E_K)^\alpha \Bigl[ \log E_K - \textstyle {1 \over \alpha} \Bigr]
\right\} \ \ \mbox{\sl (counterterms)} \nonumber
\eea
with parity preserved: $k,\ K$ in (\ref{EML}) are even for $D^+$, odd for $D^-$.

Finally, $\log D^\pm$ have (term-by-term computable) $\l \to +\infty$ expansions
called {\sl generalised Stirling expansions\/} \cite{JL}, which get constrained 
by the zeta-regularisation to a very specific form:
\bea
\label{GSE}
\log D^\pm (\l ) \si\sim\se \sum_{\alpha \in \mathcal A} a_\alpha ^\pm \bigl\langle \l ^{\alpha} \bigr\rangle 
\qquad \mbox{over the set } {\mathcal A} \mbox{ of } (\ref{SCE}) , \mbox{ with } \\
\label{CAN}
\bigl\langle \l ^{\alpha} \bigr\rangle \si\equiv\se \l ^{\alpha}
\mbox{ save for } \alpha \in {\mathbb N} : \quad
\bigl\langle \l ^0 \bigr\rangle \equiv \log \l ,
\ \bigl\langle \l ^1 \bigr\rangle \equiv \l (\log \l -1) , \ \ldots \qquad
\eea
($\ldots \bigl\langle \l ^n \bigr\rangle \equiv \l ^n(\log \l - H_n)$, $H_n =$ the harmonic numbers),
but only one implication will matter here: that {\sl no stand-alone powers\/} $b_n \l ^n$ ($n \in \mathbb N$),
including {\sl additive constants\/} ($b_0 \l ^0$), can appear.
We dub the restricted form (\ref{GSE}) ``canonical";
it actually stems more directly from a (more general) heat-trace expansion
than from (\ref{SCE}) \cite[eqs.~(8)]{V7}.

At $N=2$ (the harmonic-oscillator case),
classic $\G $-function results are recovered.
Either spectrum, say ${\mathcal E}$, essentially amounts to the set $\mathbb N$, 
for which $\mu =1$, the generalised zeta function is Hurwitz's 
$\zeta (s,\l )=\sum_{k=0}^\infty (k+\l )^{-s}$,
and the zeta-regularised determinant is
$\e^{- \partial_s \zeta (0,\l )} \equiv \sqrt{2\pi} / \Gamma(\l )$.
\cite[vol.~I eq.~1.10(10)]{EB}
Then (\ref{EML}), resp. (\ref{GSE}), restore the {\sl Euler limit formula\/}, 
resp. the {\sl Stirling expansion\/}, for
$ -\log \G (\l )$.
\cite[vol.~I eqs.~1.1(2), resp. 1.18(1)]{EB}

\subsection{Wave asymptotics}

\subsubsection{Known WKB results}

The differential equation (\ref{SSE}) on the half-line $[0,+\infty)$ has its solutions $\psi $ 
asymptotically described (for $q \to +\infty$) by {\sl WKB formulae\/},
\beq
\label{WKB}
\psi (q) \propto \Pi_\l (q)^{-1/2} 
\, \exp \Biggl\{ \pm \! \int^q \Pi_\l (\tilde q) \d \tilde q \Biggr\} , \qquad
\Pi_\l (q) \!\defi\! (V(q) \!+\! \l )^{1/2} \, ;
\eeq
i.e., $\mi \Pi_\l (q)$ is the {\sl momentum\/} in the classically forbidden region;
primitives (as in the exponent) are the {\sl action\/} functions;
$\Pi_\l (q)>0$ for $q \to +\infty$ fixes the branch.
Differentiations of (\ref{WKB}) (with respect to $q,\ \l $) are also valid.

An explicit $q \to +\infty$ form for $\psi $ stems from the classical expansion
\beq
\label{ASV}
[V(q) +  \l ]^{-s+ 1/2} \sim
\sum_\sigma \beta_\sigma (s;\vec v,\l ) \, q^{\,\sigma-Ns} , \quad
\textstyle \sigma=\frac{N}{2} ,\ \frac{N}{2} \!-\! 1,\ \frac{N}{2}  \!-\! 2,\ \ldots
\eeq
(in which the $\beta_\sigma$ come out as polynomials in $\vec v$ and $s$), as
\beq
\label{ASQ}
\psi (q) \propto q^{-N/4 \, \pm \beta_{-1}(\vec v)}
\exp \, \Biggl\{ \pm\!
{\displaystyle \sum_{\{\sigma>0\}}} \!\beta_{\sigma-1}(\vec v)
\,{\textstyle q^{\sigma} \over \textstyle \sigma} \Biggr\} 
\eeq
\cite[Thm ~6.1(ii)]{S} where $\beta_{\sigma-1} (\vec v) \defi \beta_{\sigma-1} (0;\vec v,\l )$
is independent of $\l $ for $\sigma \ge 0$ when $N \ne 2$.

\subsubsection{``Classical zeta-regularisation"}

The overall constant factor implied in (\ref{ASQ}) depends on
the primitive of $\pm\Pi_\l (q)$ selected within (\ref{WKB}). 
Against common practice and for later simplicity,
here we opt for an ``improper action", formally meant as \cite{V7}
\beq
\label{FIA}
I_\l (q) = \mbox{``} \int_q^{+\infty} \Pi_\l (\tilde q) \,\d \tilde q \mbox{ "} \qquad
\mbox{(a highly divergent integral)}.
\eeq
A natural regularisation for (\ref{FIA}) would be the analytical continuation of 
\beq
\label{PRE}
I_\l (q,s) \defi \int_q^{+\infty} (V(\tilde q) + \l )^{1/2 \, -s} \,\d \tilde q \qquad (\Re s >\mu )
\eeq
to $s=0$, but the singular structure of $I_\l (q,s)$, which follows from the expansion (\ref{ASV}) term by term,
shows (for even $N$) a generic {\sl simple pole at\/} $s=0$ of residue $ N^{-1} \beta_{-1}(\vec v) $;
just for definiteness, $\beta_{-1}(\vec v)$ is the (finite) sum \cite{V5}
\beq
\label{BET}
\beta_{-1}(\vec v) = \! \!\sum_{\{ r_j \ge 0 \}} \! \delta _{ \sum_{j=1}^{N-1} jr_j \, ,\, 1+N/2 } 
\, \frac{\G(\frac{3}{2})}{\G \bigl( \frac{3}{2} - \sum_{j=1}^{N-1} r_j \bigr) } \, 
{ v_1^{r_1} \cdots v_{N-1}^{r_{N-1}} \over r_1 ! \cdots r_{N-1} !} 
\quad ( \mbox{for } N \ne 2) .
\eeq
Thus (even though the Kronecker $\delta$ in (\ref{BET}) can often suppress the whole sum),
a fully general specification of $I_\l (q)$ must accommodate $\beta_{-1}(\vec v) \ne 0$; 
this we pursue again at $q=0$ and for an even $V(q)$ (still for simplicity only).

We then argue that (\ref{FIA}) (at $q=0$) can be fully defined in complete classical parallelism
with the zeta-regularisation of (\ref{FQD}) to $\log D(\l )$ in the quantal setting,
when $D(\l ) = \det(\hat H+\l )$ over the whole real line
or $D(\l ) = D^+(\l )D^-(\l )$, all under $V(q)$ even. 
Indeed, the classical limits of the Schr\"odinger operator $\hat H$ and of the trace operation Tr
are the Hamiltonian function $p^2+V(q)$ and the phase-space integration $(2\pi)^{-1}\int \d p \d q$
respectively. Hence the classical analogue of the zeta-function 
$Z(s,\l ) = \Tr (\hat H+\l )^{-s}$ has to be
\bea
Z_{\rm cl} (s,\l ) \si=\se 
\frac{1}{2\pi} \int_{{\mathbb R}^2} \!\! \d q \d p \, (p^2+V(q)+\l )^{-s} 
\qquad \qquad \qquad \qquad \qquad (\Re s>\mu ) \nonumber\\
\label{CZF}
\si=\se \frac{\Gamma(s \!-\! \hf)}{\sqrt \pi \,\Gamma(s)}
\! \int_0^{+\infty} \!\! (V(q) \!+\! \l )^{1/2 \, -s} \d q \equiv 
\frac{\Gamma(s \!-\! \hf)}{\sqrt \pi \,\Gamma(s)} \, I_\l (q=0,s) ,
\eea
and the last expression yields a regular continuation at $s=0$
(where the simple pole of $I_\l (q,s)$ gets cancelled by $\G (s)^{-1}$). We can thus define
\beq
\log D_{\rm cl}(\l ) \defi -\partial_s Z_{\rm cl}(s,\l )_{s=0} .
\eeq
Now when $I_\l $ is regular at $s=0$, $-\hf \partial_s Z_{\rm cl}(s,\l )_{s=0}$ evaluated through (\ref{CZF}) 
coincides with $I_\l (0,0)$ ($= \int_0^{+\infty} \Pi _\l (q) \, \d q$ formally). Inversely then,
\beq
\label{SIA}
\int_0^{+\infty} \Pi _\l (q) \, \d q \defi \hf \log D_{\rm cl}(\l ) \quad \mbox{(always finite)}
\eeq
extends the improper action to the singular-$I_\l $ case as well. 
Equivalently: in parallel to the quantal case 
(although $D_{\rm cl}$ has a branch cut instead of being entire)
$\log D_{\rm cl}$ has a {\sl canonical\/} large-$\l $ expansion;
now only one primitive of $\int_0^{+\infty} \partial_\l [\Pi _\l (q)] \, \d q $ 
can have {\sl that\/} property; {\sl it\/} then specifies $\int_0^{+\infty} \Pi _\l (q) \, \d q$.

Improper actions benefit WKB theory in many other ways, \cite{VK}
and they are no less accessible than conventional action integrals: e.g.,
\bea
\label{AVBH}
\int_0^{+\infty} \! (q^N + \l )^{1/2} \,\d q \si = \se 
- \frac{\textstyle \G (1+\frac{1}{N})\G (-\frac{1}{2}-\frac{1}{N})}{2 \sqrt \pi}
\, \l ^{ \frac{1}{2}+\frac{1}{N} } \qquad (N \ne 2) \qquad \\
\label{AVB2}
\si = \se -\qt \, \l (\log \l -1) \qquad \qquad \qquad \qquad (N =2) \\
\int_0^{+\infty} \!\! (q^4+vq^2+\l )^{1/2} \, \d q \si = \se  \qquad \qquad \qquad \qquad \qquad \qquad \qquad 
(\mbox{case } v,\ \l \ge 0) \nonumber
\eea
\beq
\label{AVT}
\left\{ \matrix{
\frac{1}{3} (v + 2\sqrt \l )^{1/2} \bigl[ 2 \sqrt \l K(k) - v E(k) \bigr] , &
k \defi \Bigl( \frac{\textstyle v - 2 \sqrt \l}{\textstyle v + 2 \sqrt \l} \Bigr) ^{1/2} && 
(v \ge 2\sqrt \l ) \cr
\frac{1}{3} \, \l ^{1/4} \bigl[ (2\sqrt \l + v) K(\tilde k) -2 v E(\tilde k) \bigr] , \hfill &
\tilde k \defi \frac{\textstyle (2\sqrt \l - v)^{1/2}}{\textstyle 2 \, \l ^{1/4} } \hfill &&
(v \le 2\sqrt \l ) } \right.
\eeq
where $E(\cdot),\ K(\cdot)$ are the usual complete elliptic integrals. 
\cite[Vol.~II, \S 13.8]{EB}

The calculation as above but for the ratio $D^+/D^-$ in place of the product $D^+D^-$ is now regular,
and its output, $[D^+_{\rm cl} / D^-_{\rm cl}] (\l ) \equiv \Pi _\l (0)$,  
completes the results we need to rewrite the WKB formulae suggestively for later analogy.
Together with (\ref{SIA}), and with translation covariance restored
(thanks to $\int_{\mathcal I} \Pi_\l (\tilde q) \, \d \tilde q$ being canonical 
for finite intervals $\mathcal I$), 
all of that converts the canonically normalised recessive WKB-solution (\ref{WKB}) 
plus its $q$-derivative partner, namely
\beq
\label{CSC}
\left. \matrix
{\hfill \psi _\WKB (q) = +\Pi_\l (q)^{-1/2} \, \exp \!\int_q^{+\infty} \Pi_\l (\tilde q) \, \d \tilde q , \cr
[\psi ']_\WKB (q) = -\Pi_\l (q)^{+1/2} \, \exp \!\int_q^{+\infty \strut} \Pi_\l (\tilde q) \, \d \tilde q , 
} \right\}
\eeq
to the form (also valid for non-even $V$)
\beq
\label{SCI}
\psi_\WKB (q) \equiv D_{q,\rm cl}^- (\l ), \qquad \qquad [\psi']_\WKB (q) \equiv -D_{q,\rm cl}^+ (\l ),
\eeq
where $D_{q, \rm cl}^\pm (\l )$ are the ``classical determinants" defined like $D_{\rm cl}^\pm (\l )$
but over the half-line $[q,+\infty)$ in place of $[0,+\infty)$.

\section{Exact analysis of the Schr\"odinger equation}

The differential equation (\ref{SSE}) is known to
admit exact solutions $\psi (q)$ which are {\sl recessive\/}, i.e., decaying for $q \to +\infty$,
according to the asymptotic WKB form (\ref{WKB}). We specifically select
the exact solution $\psi_\l (q)$ that is {\sl canonically recessive\/}: 
i.e., asymptotic to our normalised (decaying) WKB form (\ref{CSC}),
\beq
\label{PSI}
\psi_\l (q) \sim \psi _\WKB (q) \equiv \Pi_\l (q)^{-1/2} \, \exp \int_q^{+\infty} \Pi_\l (\tilde q) \d \tilde q 
\qquad \mbox{for }q \to +\infty .
\eeq

\subsection{Basic exact-WKB identities}

The exact solution $\psi_\l $, {\sl given the normalisation (\ref{PSI}) for\/} $q \to +\infty$, 
is then expressed at finite $q$ by the {\sl exact identities\/} \cite{V7}
\beq
\label{BID}
\psi_\l (q) \equiv D_q^- (\l ), \qquad  \psi'_\l (q) \equiv -D_q^+ (\l ),
\eeq
in terms of the {\sl zeta-regularised\/} Dirichlet ($-$), resp. Neumann ($+$), spectral determinants 
over the half-line $[q,+\infty)$.

The crucial feature of (\ref{BID}) is the unseen one! No fudge factor whatsoever
in-between the solution $(\psi _\l , \psi'_\l )$ and the spectral determinants $D^\pm$,
thanks to careful normalisations, is what will allow for exact solution algorithms.

In view of the alternative form (\ref{SCI}) exhibited by the usual asymptotic-WKB solutions,
we see (\ref{BID}) as {\sl exact-WKB formulae\/} for the exact solution~$\psi _\l $.
Clearly these do not appear as explicit as the asymptotic-WKB solutions,
expressible by quadratures as in (\ref{CSC}).
Nevertheless, our main point is this: (\ref{BID}), too, will end up ``solving" 
the Schr\"odinger problem analytically, and in a more roundabout but now exact, manner.

Proof of (\ref{BID}) (adapted from \cite[Apps. A \& D]{V1}). It has two steps: 
1) prove the first logarithmic derivatives of the identities (\ref{BID}) with respect to $\l $; 
2) do one controlled integration thereupon to attain (the logarithm of) (\ref{BID}) itself.
(All that for $N>2$; here we skip the particular cases $N=1$ and 2, which need two differentiations/integrations.)
We denote $\dot{} \equiv \frac{\d}{\d \l }$ and, without loss of generality, set $q=0$ again.

1) Applying to (\ref{SQD}) the basic functional relation (\ref{ZFR}) tuned to ${s=0}$,
$[ \partial_s Z(s,\l )_{s=0} ] \, \dot {} = -Z(1,\l ) $,
we find the logarithmic derivative of (\ref{BID}) with respect to $\l $, at $q=0$, to be
\beq
\label{BIP}
\dot \psi _\l (0) / \psi _\l (0) \equiv Z^-(1,\l ) , \qquad
\dot \psi '_\l (0) / \psi '_\l (0) \equiv Z^+(1,\l ) .
\eeq
We then refer to the Appendix for a {\sl trace-formula\/} proof of (\ref{BIP}).

2) One integration upon (\ref{BIP}) with respect to $\l $ gives
\beq
\log \psi_\l (0) \equiv \log D^- (\l ) + C^-, \qquad \log [-\psi'_\l (0)] \equiv \log D^+ (\l ) + C^+,
\eeq
where $C^\pm$ are integration constants. 
(For sign-fixing: every argument of $\log(\cdot)$ has to be positive for large $\l >0$.)
Now, WKB asymptotics like (\ref{PSI}) also hold for $\l \to +\infty$ at fixed $q$,
and we have painstakingly normalised earlier to ensure {\sl canonical\/} large-$\l $ behaviours on all sides; 
therefore, any extra additive constants like $C^\pm$ are banned by (\ref{CAN}), 
and (\ref{BID}) results.
$\square$

\subsection{Explicit Wronskian identity}

A second solution of (\ref{SSE}) besides $\psi _\l $, 
now recessive for $q \to + \e^{-\mi\varphi/2} \!\infty$,
can be specified through the {\sl first conjugate\/} equation, 
(\ref{CJE}) for $\ell=1$, \cite[\S 2.7]{S} as
\beq
\label{FCE}
\Psi_\l (q) \defi \psi_{\l ^{[1]}}^{[1]} (\e^{\mi\varphi/2}q) .
\eeq
Because the asymptotic form (\ref{ASQ}) for $\psi _\l $ and its analogues for $\Psi_\l (q)$ 
and the $q$-derivatives all hold over the positive direction,
the constant Wronskian of $\Psi _\l $ and $\psi _\l $, 
$W(\Psi _\l , \psi _\l ) \defi \Psi_\l \psi'_\l -\Psi'_\l \psi_\l $,
can be deduced {\sl explicitly\/}, just from their $q \to +\infty$ asymptotics. \cite[Thm~21.1(iii)]{S} 
Under the canonical normalisations and recalling (\ref{BET}) for $\beta_{-1}(\vec v)$, it is \cite[including Corrigendum]{V5}
\beq
\label{EWR}
W(\Psi _\l , \psi _\l ) \equiv 2\,\mi\e^{\mi\varphi/4} \e^{\mi\varphi\beta_{-1}(\vec v)/2} \qquad (N \ne 2).
\eeq

This basic {\sl Wronskian identity\/} is the other essential piece of the framework: 
if we now use the exact-WKB identities (\ref{BID}) within the alternative form 
$W(\Psi _\l , \psi _\l ) = \Psi_\l (0)\psi'_\l (0)-\Psi'_\l (0)\psi_\l (0)$,
this then translates (\ref{EWR}) into a {\sl functional relation between spectral determinants\/},
which for $N \ne 2$ reads
\beq
\label{FFR}
\e^{+\mi\varphi/4} D^{[1]+} (\e^{-\mi\varphi} \l ) D^-(\l )
-\e^{-\mi\varphi/4} D^+(\l ) D^{[1]-} (\e^{-\mi\varphi} \l ) 
\equiv 2 \,\mi \e^{\mi\varphi\beta_{-1}(\vec v) /2} .
\eeq

\subsection{Exact quantisation conditions}

At first glance, the original Wronskian identity (\ref{EWR}), or its sibling (\ref{FFR}), 
look grossly underdetermined: one equation for two unknown functions. 
This view however proves a delusion here, thanks to the structure formulae (\ref{EML}). 
Their first effect is to reduce quests for determinants
to quests for their zeros (the eigenvalues): 
in other words, we are to seek {\sl exact quantisation conditions\/}.
The latter problem then becomes central to the whole framework; we now sketch our solution for it (as valid for $N>2$). \cite{V7}

To analyse the eigenvalues $\{E_k\}$ of the Schr\"odinger operator $\hat H$ on the half-line, 
we may divide the functional relation (\ref{FFR}) at $\l =-E_k$ by its first conjugate partner, to get
\beq
\label{EQ1}
2 \arg  D^{[1]\pm}(-\e^{-\mi\varphi} E) -\varphi \, \beta_{-1}(\vec v) \Bigm| _{E=E_k}
= \pi [k+\hf \pm \textstyle \frac{N-2}{2(N + 2)}] \qquad
{\rm for\ } k \ {\rm even \atop odd} \ .
\eeq
The formulae (\ref{EQ1}) have the outer form of {\sl Bohr--Sommerfeld quantisation conditions\/}
(smooth left-hand-side functions of $E$, sampled at integers).
Now they are {\sl exact, albeit not closed\/}
(the left-hand sides invoke the determinants of the first conjugate spectra, equally unknown).

However, in either (Neumann / Dirichlet) sector --- there is a complete decoupling between them ---,
the set of {\sl all\/} quantisation conditions for the $L$ conjugate problems
plus the $L$ structure formulae (now used to reconstruct each determinant from its spectrum),
add up to a {\sl formally closed\/} system:
\bea
\label{EQC}
\textstyle \mi^{-1}
\Bigl[ \log D^{[\ell+1]\pm} (-\e^{-\mi\varphi}E_k^{[\ell]})
- \log D^{[\ell-1]\pm} (-\e^{+\mi\varphi}E_k^{[\ell]}) \Bigr]
\si-\se (-1)^\ell \varphi \beta_{-1}(\vec v) \\[4pt]
\si=\se
\pi [ k+\hf \pm \textstyle \frac{N-2}{2(N+2)} ] \, , \nonumber
\eea
\bea
\label{CSE}
\log D^{[\ell]\pm} (\l ) \si\equiv\se \lim_{K \to +\infty} \Biggl \{
\sum_ {k<K} \log (E_k^{[\ell]} + \l )
+ \hf \log (E_K^{[\ell]} + \l ) \Biggr . \\[-10pt]
&& \qquad \qquad \qquad \qquad \qquad - \hf \sum_{ \{\alpha >0 \} } \Biggl.
b_\alpha^{[\ell]} ( E_K^{[\ell]} ) ^\alpha
[ \log E_K^{[\ell]} - \textstyle {1 \over \alpha} ] \Biggr \}  \nonumber
\eea
for $\ell \in {\mathbb Z} / L {\mathbb Z}$ and $k,K \ {\rm even \atop odd}$.
The selfconsistency brought by (\ref{CSE}) makes this
a {\sl fixed-point problem\/}, with parities decoupled, hence abstractly of the form
${\mathcal M}^\pm \{ {\mathcal E}^{\bullet \pm} \}= {\mathcal E}^{\bullet \pm}$, 
where ${\mathcal E}^{\bullet \pm}$ (``compound spectra") denote the disjoint unions of
all the $L$ conjugate spectra ${\mathcal E}^{[\ell] \pm}$ of a given parity (Fig.~1),
and ${\mathcal M}^\pm $ are some spectra-to-spectra mappings.
Note:  as a completely general fact, {\sl concrete\/} such forms and mappings are highly non-unique;
we just hope to find {\sl some\/} explicit such form(s) of (\ref{EQC})--(\ref{CSE}) that behave nicely.

The equations (\ref{CSE}) furthermore require {\sl asymptotically correct\/} arguments 
${\mathcal E}^{\bullet \pm}$, for the $K \to +\infty$ convergence to take place.

\begin{figure}[h]
\center
\includegraphics[scale=.62,angle=-90]{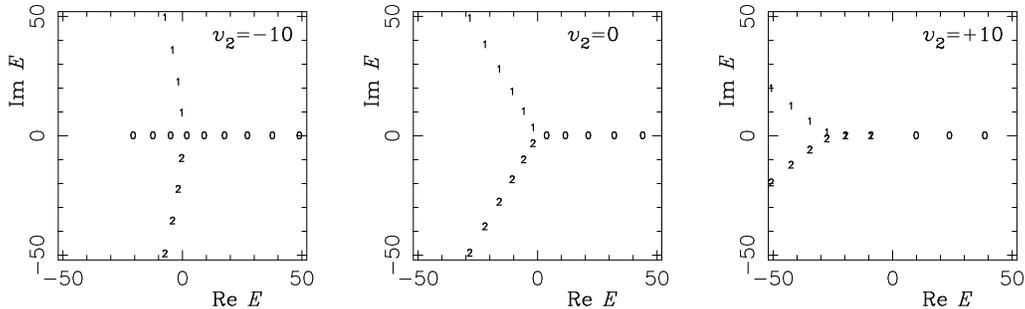}
\caption{\label{F1} \small (from \cite{V5}) 
The (odd) ``compound spectrum" ${\mathcal E}^{\bullet -}$ for quartic potentials 
$V(q) = q^4+v_2q^2$, at $v_2=-10$ (left), 0 (center), $+10$ (right). 
(In each square the $L=3$ sets ${\mathcal E}^{[\ell]-}$ are seen finitely cropped,
labelled by their respective $\ell$-values 0, 1, 2,
and rotated to fit their relative positions in the system (\ref{EQC}).)}
\end{figure}

In the simplest case, all homogeneous potentials $V(q)=q^N$: \cite{V2,DT}
the compound spectra amount to the original ones ${\mathcal E}^\pm \subset {\mathbb R}_+$,
mappings ${\mathcal M}^\pm$ exist in the real form: $\{ E'_k >0\} \mapsto \{ E''_k >0\}$
($k \ {\rm even \atop odd}$), given by the formulae
\beq
\label{EQN}
\matrix{\frac{2}{\pi} \, A_\pm (E''_k) =& k+\hf \pm \frac{N-2}{2(N + 2)} \hfill
& {\rm for\ } k \ {\rm even \atop odd} \ , \quad {\rm where} \hfill \cr
\hfill A_\pm (E'') \defi& \sum\limits_\ell \arg (E'_\ell - \e^{-\mi \varphi} E'') 
& {\rm for\ } \ell \ {\rm even \atop odd} \ , \ E''>0, \quad (N \ne 2),}
\eeq
and these have now been {\sl proven globally contracting\/}. \cite{A}
Next, for moderately inhomogeneous potentials:
we have only tested quartic and sextic cases numerically, \cite[\S 3]{V5}
and still -- but now without proof to date -- observed contracting maps.
As long as this holds, the unknowns in (\ref{EQC})--(\ref{CSE}) 
(the eigenvalues and the determinants)
get {\sl uniquely\/} determined by the fixed-point equations, 
and {\sl effectively\/} reached as limits under iterations of the contracting maps;
thus the fixed-point equations qualify as (exact) quantisation conditions.
Only in {\sl some\/} larger-$\vec v$ regions do our {\sl current\/} schemes become {\sl numerically\/} unstable
-- a non-elucidated issue. (Note: all numerical mappings have to be
finite-dimensional as well, which entails {\sl some approximation\/}.)

Again for homogeneous potentials, moreover, (\ref{EQN}) mysteriously coincide 
with the Bethe-Ansatz equations for certain solvable models of 2D statistical mechanics.
This has fostered a new research area, the {\sl ODE/IM correspondence\/},
relating our exact-WKB solvability with complete integrability in the modern sense. 
\cite{DT,BLZ,SU,DDT}

We may then briefly list 1D Bohr--Sommerfeld quantisation conditions for the eigenvalues $E_k$,
in the common general form $A(E_k) = k+\hf$, as:

\ni - old exact-WKB setting, special: the $E_k$ get specified exactly via an explicit function $A(E)$,
but only for exceptional systems (e.g., harmonic oscillator);

\ni - old general WKB approaches, semiclassical: the $E_k$ get specified 
via an explicit function or expansion $A(E)$, but only asymptotically for $k \to +\infty$;

\ni - our general exact-WKB formalism: the $E_k$ get specified exactly via a function $A(E)$ 
itself specified exactly, but implicitly through a fixed-point equation which,
while infinite-dimensional, is explicit.

\subsection{Solving for the unknown functions $\psi$}

\begin{figure}[h]
\center
\includegraphics[scale=.5]{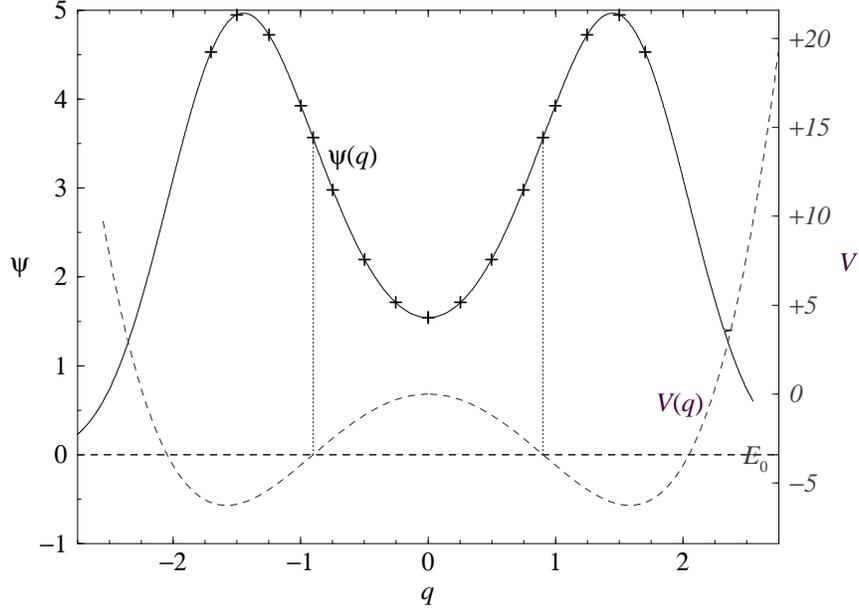}
\caption{\label{F2} \small (from \cite{V8}) 
Exact-WKB calculations for an eigenfunction $\psi(q)$ (the ground state in a double-well potential).
{\bf Dashed lines} (with right-hand scale): for reference,
the potential $V(q)=q^4-5q^2$ and its lowest eigenvalue $E_0 \approx -3.41014$ (input);
vertical lines mark the two inner turning points.
{\bf Continuous line} (with left-hand scale): the eigenfunction $\psi $, 
canonically normalized as (\ref{PSI}),
drawn by numerical integration of (\ref{SSE}) (NAG routine D02KEF).
{\large \tt +}~: exact-WKB values $\psi(q)$ according to (\ref{BID}): 
meaning, at every $q$, a numerical iteration limit of
the fixed-point system (\ref{EQC})--(\ref{CSE}) for the determinant value $D_q^- (-E_0)$.
The plot validates the exact-WKB method in extreme quantal regimes: a ground state, turning points
(on the other hand, an unexplained numerical instability sets in beyond $|q| \approx 1.7$).}
\end{figure}

We focus on finding the canonical recessive solution $\psi _\l $ for {\sl arbitrary\/} $\l $
(once a particular solution is known, others trivially follow).

Equations (\ref{BID}) have specified $\psi _\l (q)$ (and $\psi '_\l (q)$) as particular spectral determinants
over the half-line $[q,+\infty)$; now the latter are of the type fully resolved 
by the exact quantisation conditions above. Consequently, the unknown wavefunctions $\psi $
are computable as well in the exact approach (as numerical tests confirm: Fig.~2). \cite{V5,V8}

\subsection{Toward 1D quantum perturbation theory}

On potentials like $V(q)=q^2+gq^4$, it is possible to monitor the $g \to 0^+$ limit
of the exact-WKB framework, in spite of its extremely singular character 
(thus the degree $N$, which is the most critical parameter in the framework, jumps from 4 to 2 at $g=0$).
Results are largely governed by the singular behaviours of improper action integrals,
like ${\int_0^\infty (q^4+g^{-2/3}q^2+\l )^{1/2} \, \d q}$ \ ($g \to 0^+$) for the case $V(q)=q^2+gq^4$. 
\cite{V9,VK}

\begin{appendix}

\section*{Appendix: derivation of (\ref{BIP}) as trace formulae}

The identities (\ref{BIP}) awaiting proof indeed take the form of {\sl trace formulae\/},
\beq
\label{TF}
\dot \psi _\l (0) / \psi _\l (0) \equiv \Tr (\hat H+\l )_-^{-1}\, , \qquad
\dot \psi '_\l (0) / \psi '_\l (0) \equiv \Tr (\hat H+\l )_+^{-1}\, ,
\eeq
where $\psi _\l $ is the solution (\ref{PSI}) of $(\hat H+\l )\psi =0$
recessive for $q \to +\infty$, 
and ${(\hat H+\l )_\pm^{-1}}$ are the resolvent operators (Green's functions) on the half-line $q \in [0,+\infty)$
with the corresponding boundary conditions at $q=0$ ($-=$ Dirichlet, $+=$ Neumann).
The resolvents are trace-class under our assumption $N>2$.

Note: the $\psi $-normalisations, so essential in the main text, inversely become irrelevant here;
already in (\ref{TF}), any recessive solution $c \psi _\l $ will do.

If now $\psi^\pm _\l (q)$ is the $\rm Neumann \atop Dirichlet$ solution of $(\hat H+\l )\psi =0$,
also unique up to a factor,
then the integral kernels $G_\l ^\pm (q,\tilde q)$ of the Green's functions are known, in this 1D setting, to be
\beq
G_\l ^\pm (q,\tilde q) = W(\psi _\l ,\psi ^\pm_\l )^{-1} 
\psi ^\pm_\l \bigl( \min(q,\tilde q) \bigr) \, \psi_\l \bigl( \max(q,\tilde q) \bigr) \quad (q,\ \tilde q \in {\mathbb R}_+) ,
\eeq
where $W(\psi ,\phi ) \defi \psi \phi ' - \phi \psi '$ denotes the Wronskian of $(\psi ,\phi )$ 
(a constant when $(\psi ,\phi )$ is a solution pair). As intermediate result, then,
\beq
\label{TRF}
\Tr (\hat H+\l )_\pm^{-1} = \int_0^\infty G_\l ^\pm (q,q) \, \d q
= W(\psi _\l ,\psi ^\pm_\l )^{-1} \int_0^\infty \psi ^\pm_\l (q) \psi_\l (q) \, \d q .
\eeq

Next, the subtraction of $\dot \psi _\l \, \bigl[ (\hat H+\l ) \psi ^\pm_\l \bigr] = 0$
from $\psi _\l^\pm \, \bigl[ (\hat H+\l ) \psi _\l \bigr] \, \dot {} = 0$ (two obvious equalities) yields
\beq
\psi ^\pm_\l \psi_\l = \bigl[ W(\psi ^\pm_\l ,\dot \psi _\l )\bigr] ' \, ;
\eeq
this reveals an explicit primitive for the product $\psi ^\pm_\l \psi_\l $, reducing (\ref{TRF}) to
\beq
\Tr (\hat H+\l )_\pm^{-1} = W(\psi _\l ,\psi ^\pm_\l )^{-1} \Bigl[ W(\psi ^\pm_\l ,\dot \psi _\l ) \Bigr] _0^\infty .
\eeq
But as $q \to +\infty$, using (\ref{WKB}) and its derivatives, $W(\psi ^\pm_\l ,\dot \psi _\l ) (q)
= {\rm O} \bigl[ \int_q^\infty \d \tilde q / \Pi_\l (\tilde q) \bigr]$ 
which tends to 0 under $N>2$, hence
\beq
\Tr (\hat H+\l )_\pm^{-1} = -W(\psi _\l ,\psi ^\pm_\l )^{-1} \, W(\psi ^\pm_\l ,\dot \psi _\l )_{q=0} .
\eeq
Finally, this collapses to the identity pair (\ref{TF}) 
when $W(\psi _\l ,\psi ^\pm_\l )$ (constant) is itself expressed at $q=0$ 
and the boundary conditions for $\psi ^-_\l $, resp. $\psi ^+_\l $, are applied. $\square$

\end{appendix}

\end{document}